\documentclass[11pt,a4paper]{article}
\usepackage{amsmath} 
\usepackage{amssymb} 
\usepackage{bm}  

\def\be{\begin{equation}}
\def\ee{\end{equation}}
\def\ba{\begin{eqnarray}}
\def\ea{\end{eqnarray}}
\def\onehalf{{\textstyle{\frac{1}{2}}}}

\begin{document}

\begin{center}
{\Large \bf de Sitter geodesics: reappraising the notion of motion}
\vskip 0.5cm
{\bf J. G. Pereira and A. C. Sampson}
\vskip 0.2cm
{\it Instituto de F\'{\i}sica Te\'orica, UNESP-Univ Estadual Paulista \\
Caixa Postal 70532-2, 01156-970 S\~ao Paulo, Brazil}
\end{center}

\vskip 0.3cm
\begin{quote}
{\bf Abstract.}~{\footnotesize The de Sitter spacetime is transitive under a combination of translations and proper conformal transformations. Its usual family of geodesics, however, does not take into account this property. As a consequence, there are points in de Sitter spacetime which cannot be joined to each other by any one of these geodesics. By taking into account the appropriate transitivity properties in the variational principle, a new family of maximizing trajectories is obtained, whose members are able to connect any two points of the de Sitter spacetime. These geodesics introduce a new notion of motion, given by a combination of translations and proper conformal transformations, which may possibly become important at very-high energies, where conformal symmetry plays a significant role.

}
\end{quote}

%

\vskip 0.6cm
\section{Introduction}

Spacetimes with constant sectional curvature are maximally symmetric in the sense that they can lodge the highest possible number of Killing vectors \cite{weinberg}. Their Riemann curvature tensor are completely specified by the scalar curvature $R$, which is constant throughout spacetime. Minkowski spacetime $M$, with vanishing scalar curvature, is the simplest one. Its kinematic group is the Poinca\-r\'e group ${\mathcal P} = {\mathcal L} \oslash {\mathcal T}$, the semi-direct product of Lorentz ${\mathcal L}$ and the translation group ${\mathcal T}$. It is a homogeneous space under the Lorentz group:
\[
M = {\mathcal P}/{\mathcal L}.
\]
The invariance of $M$ under the transformations of ${\mathcal P}$ reflects its uniformity. The Lorentz subgroup provides an isotropy around a given point of $M$, and the translation symmetry enforces this isotropy around any other point. This is the meaning of homogeneity: all points of spacetime are ultimately equivalent.

In addition to Minkowski, there are two other maximally symmetric four-dimensional spacetimes with constant sectional curvature. One is de Sitter, with topology $R^1 \times S^3$ and (let us say) negative scalar curvature. The other is anti-de Sitter, with topology $S^1 \times R^3$ and positive scalar curvature. As hyperbolic spaces, though, both have negative Gaussian curvature. Our interest here will be concentrated on the de Sitter spacetime, denoted $dS(4,1)$, whose kinematic group is the de Sitter group $SO(4,1)$. Like Minkowski, it is a homogeneous space under the Lorentz group:
\[
dS(4,1) = SO(4,1) / {\mathcal L}.
\]
Of course, in order to be physically relevant, it must be solution to Einstein equation. However, similarly to Minkowski, it is more fundamental than Einstein equation in the sense that, as a quotient space it is known {\it a priori}, independently of gravitation. In fact, observe that the de Sitter metric does not depend on the gravitational constant.

As fundamental spacetimes, Minkowski and de Sitter represent different backgrounds for the construction of any physical theory. General relativity, for instance, can be construct on either of them. Of course, the two theories will have the same dynamics; only their local kinematics will be different. This means that only the strong equivalence principle will change. In fact, the strong equivalence principle states that it is always possible to find a local frame in which the laws of physics reduce to those of special relativity. If the underlying local spacetime is Minkowski, the kinematics will be ruled by the Poincar\'e group. If the underlying local spacetime is de Sitter, the kinematics will be ruled by the de Sitter group \cite{dSsr0,dSsr1,dSsr2}.

Although the isotropy of both Minkowski and de Sitter is determined by the Lorentz group, their homogeneity properties are completely different: whereas Minkowski is transitive under spacetime translations, de Sitter is transitive under a combination of translations and proper conformal transformations \cite{dSsr0}. The de Sitter group, therefore, naturally introduces the proper conformal transformations in the spacetime kinematics,\footnote{It is important to note that the inclusion of the proper conformal transformations does not change the number of degrees of freedom of the spacetime kinematics. In fact, Poincar\'e --- the kinematic group of Minkowski spacetime --- and de Sitter --- the kinematic group of the de Sitter spacetime --- are both ten-dimensional.} a property whose consequences have been mostly overlooked. For example, considering that transitivity is deeply related to the notion of motion, that is, to the way one moves from one point to another, the presence of the proper conformal transformation in the transitivity of the de Sitter spacetime has immediate implications for its geodesics. Such property, however, has never been taken into account in the variational principle used to obtain the geodesics of the de Sitter spacetime. For this reason, the geodesics usually obtained describe trajectories whose points are connected to each other by ordinary translations only. As a consequence, there are points in the de Sitter spacetime which cannot be joined to each other by any geodesic of this family \cite{ellis}. By taking into account the appropriate transitivity properties of the de Sitter spacetime, the purpose of this paper is to look for a new family of trajectories which are able to connect any two points of the de Sitter spacetime.

\section{de Sitter transformations}
\label{sec:deSitterKine}


\subsection{de Sitter spacetime}

The maximally symmetric de Sitter spacetime $dS(4,1)$ can be seen as a hyper-surface in the ``host'' pseudo-Euclidean spaces ${\bf E}^{4,1}$, inclusion whose points in Cartesian coordinates $\chi^A$ $(A, B, ... = 0, ..., 4)$ satisfy
\be
\eta_{AB} \chi^A \chi^B = - \, l^2,
\ee
where $\eta_{AB}$ = $(+1,-1,-1,-1,-1)$ and $l$ is the de Sitter length-parameter (or pseudo-radius). The four-dimensional stereographic coordinates $x^\mu$ ($\mu, \nu, ... = 0, ..., 3$) are obtained through a projection from the de Sitter hyper-surface into a target Minkowski spacetime. They are defined by \cite{gursey}
\be
\chi^\mu = \Omega(x) \, x^\mu \qquad \mbox{and} \qquad
\chi^4 = -\, l \, \Omega(x) \left(1 + \frac{\sigma^2}{4 l^2} \right),
\label{stepro}
\ee 
where
\be
\Omega(x) = \frac{1}{1 - {\sigma^2}/{4 l^2}},
\label{n}
\ee
with $\sigma^2$ is the Lorentz-invariant quadratic form $\sigma^2 = \eta_{\mu \nu} \, x^\mu x^\nu$. In terms of the stereographic coordinates, the metric of the de Sitter spacetime assumes the conformally flat form
\be
g_{\mu \nu} = \Omega^2(x) \, \eta_{\mu \nu}.
\label{44}
\ee
The Christoffel connection of this metric is \cite{livro}
\be
\Gamma^{\lambda}{}_{\mu \nu} = \left( \delta^{\lambda}{}_{\mu}
\delta^{\sigma}{}_{\nu}  + \delta^{\lambda}{}_{\nu}
\delta^{\sigma}{}_{\mu} - \eta_{\mu \nu} \eta^{\lambda \sigma} \right)
\partial_\sigma \ln \Omega.
\label{46}
\ee
The corresponding Riemann tensor components are found to be
\be
R^{\mu}{}_{\nu \rho \sigma} = - \, \frac{1}{l^2}
\left(\delta^{\mu}{}_{\rho} g_{\nu \sigma} - \delta^{\mu}{}_{\sigma} g_{\nu
\rho} \right).
\label{47}
\ee

\subsection{de Sitter generators and homogeneity}

An infinitesimal de Sitter transformation is written as
\be
\delta \chi^A = \onehalf \, \epsilon^{CD} L_{CD} \, \chi^A
\ee
where $\epsilon^{CD} = - \epsilon^{DC}$ are the transformation parameters, and
\be
L_{CD} = \eta_{CE} \, \chi^E \, P_D - \eta_{DE} \, \chi^E \, P_C,
\ee
with $P_C = \partial/\partial \chi^C$, are the de Sitter generators. In terms of stereographic coordinates, it assumes the form
\be\label{ds-t}
\delta x^\mu \equiv \delta_L x^\mu + \delta_{\Pi} x^\mu =
\onehalf \epsilon^{\rho \sigma} L_{\rho \sigma} x^\mu +
\epsilon^\rho \Pi_\rho x^\mu,
\ee
where $\epsilon^{\rho \sigma}$ and $\epsilon^{\rho} \equiv l \, \epsilon^{4 \rho}$ are the new transformation parameters,
\be\label{lorentzdS}
L_{\rho \sigma} =
\eta_{\rho \lambda} \, x^\lambda \, P_\sigma - \eta_{\sigma \lambda} \, x^\lambda \, P_\rho
\ee
are the Lorentz generators, and
\be\label{pi}
\Pi_\rho \equiv \frac{L_{4 \rho}}{l} =
P_\rho - \frac{1}{4 l^2} \, K_\rho
\ee
are the so-called de Sitter ``translation'' generators, with
\be
P_\rho = \partial/ \partial x^\rho \quad \mbox{and} \quad
K_\rho = \left(2 \eta_{\rho \nu} \, x^\nu x^\mu - \sigma^2 \delta_{\rho}{}^{\mu} \right) 
\partial/ \partial x^\mu,
\label{TransGenerators}
\ee
respectively, the translation and the proper conformal generators. The reason for the quotation is related to the fact that they are not actually translations, but rotations, as can be seen from their commutation relation \cite{zimer}:
\be
[{\Pi}_{\rho}, {\Pi}_{\mu}] = - \, l^{-2} \,
{L}_{\rho \mu}.
\ee
We see from the generators (\ref{lorentzdS}) and (\ref{pi}) that, from the algebraic point of view, the only difference between Minkowski and de Sitter spacetimes is the notion of uniformity. In Minkowski, the transitivity is defined by ordinary translations. In de Sitter spacetime, on the other hand, it is defined by a combination of translations and proper conformal transformations.

\subsection{Killing vectors}
\label{Killing}

Substituting the generators (\ref{lorentzdS}) and (\ref{pi}) in the transformation (\ref{ds-t}), the Lorentz transformation acquires the form
\be
\delta_{L} x^\mu = \onehalf \, \xi^{\;\mu}_{\rho \sigma} \, \epsilon^{\rho \sigma},
\ee
where
\be
\xi^{\;\mu}_{\rho \sigma} = \left( \eta_{\rho \lambda} \, \delta^\mu_\sigma - 
\eta_{\sigma \lambda} \, \delta^\mu_\rho \right) x^\lambda
\ee
are the Killing vectors of the Lorentz group. The de Sitter ``translations'', on the other hand, becomes
\be
\delta_{\Pi} x^\mu = \xi^{\mu}_{\rho} \, \epsilon^{\rho},
\label{dStrans}
\ee
where
\be
\xi_{\rho}^\mu = 
\delta_\rho^\mu - \frac{1}{4l^2} \, \bar{\delta}_\rho^\mu
\label{dsKilling}
\ee
are the corresponding Killing vectors, with $\delta_\rho^\mu$ the Killing vector of translations, and
\be
\bar{\delta}_\rho^\mu = 2 \eta_{\rho \nu} \, x^\nu x^\mu -
\sigma^2 \delta_\rho^\mu
\ee
the Killing vectors of proper conformal transformations.

\section{Conservation laws and geodesics}

Let us consider now a general matter field with Lagrangian density ${\mathcal L}_m$. Its action integral is
\be
S_m = \frac{1}{c} \int {\mathcal L}_m \, \sqrt{-g} \, d^4x.
\ee
The invariance of this action under the de Sitter translation (\ref{dStrans}) yields the covariant conservation law \cite{paper2}
\be
\nabla_\mu \Pi^{\rho \mu} = 0,
\label{dSconservation}
\ee
where
\be
\Pi^{\rho \mu} \equiv \xi^{\mu}_{\lambda} \, T^{\rho \lambda} =
T^{\rho \mu} - \frac{1}{4l^2} \, {K}^{\rho \mu}
\label{TmK}
\ee
is the Noether current, with
\be
T^{\rho \mu} = -\, \frac{2}{\sqrt{-g}} \, \frac{\delta {\mathcal L}_m}{\delta g_{\rho \mu}}
\label{syem}
\ee
the symmetric energy-momentum current, and
\be
K^{\rho \mu} =
\left(2 \eta_{\lambda \nu} \, x^\nu x^\mu -
\sigma^2 \delta_\lambda{}^\mu \right) T^{\rho \lambda}
\label{KdelT}
\ee
the proper conformal current~\cite{coleman}. Instead of the energy-momentum current, what is conserved now is a combination of energy-momentum and proper conformal currents. This is crucial point: only in de Sitter spacetime, or in any other curved spacetime that reduces locally to de Sitter, the proper conformal current appears as part of the conserved current. In Minkowski, or in any other curved spacetime that reduces locally to Minkowski, the conserved quantity includes the energy-momentum current only. This shows how the change in the spacetime uniformity produces concomitant changes in the conserved currents.

In general, neither $T^{\mu \nu}$ nor $K^{\mu \nu}$ is conserved separately. In fact, as an explicit calculation shows, they satisfy
\be
\nabla_\mu T^{\mu \nu} = \frac{2 \, T^\rho{}_\rho \, x^\nu}{4l^2 - \sigma^2} \qquad \mbox{and} \qquad
\nabla_\mu {K}^{\mu \nu} = \frac{2 \, T^\rho{}_\rho \, x^\nu}{1 - \sigma^2/4l^2}.
\label{Conser1}
\ee
In the contraction limit $l \to \infty$, de Sitter reduces to Minkowski spacetime, and we obtain the usual conservation laws \cite{coleman2}
\be
\partial_\mu T^{\mu \nu} = 0 \qquad \mbox{and} \qquad
\partial_\mu {K}^{\mu \nu} = 2 \, T^\rho{}_\rho \, x^\nu.
\ee
In this limit, physics becomes invariant under spacetime translations, and the energy-momen\-tum current turns out to be conserved.


Integrating the covariant conservation law (\ref{dSconservation}) in a space section of the de Sitter spacetime, and using the single-pole approximation \cite{mathi,papa,desabata}, we obtain the momentum conservation law,
\be
\frac{d \pi^\mu}{ds} +
\Gamma^\mu{}_{\nu \lambda} \, \pi^\nu \, u^\lambda = 0, 
\label{GendSGeod0}
\ee
where $\Gamma^\mu{}_{\nu \lambda}$ is the Christoffel connection (\ref{46}), and
\be
\pi^\mu = \xi^{\mu}_{\rho} \, p^\rho \equiv p^\mu - \frac{1}{4l^2} \, k^\mu,
\label{dSmome}
\ee
is the de Sitter momentum, with
\be
p^\rho = m c u^\rho \qquad \mbox{and} \qquad k^\mu = \left(2 \eta_{\rho \nu} \, x^\nu x^\mu - \sigma^2 \delta_\rho{}^\mu \right) p^\rho,
\ee
respectively, the ordinary and the proper conformal momenta. Considering that in special relativity the conservation of (ordinary) momentum
\be
\frac{d p^\mu}{ds} = 0
\ee
defines the geodesics of Minkowski spacetime, the  conservation law (\ref{GendSGeod0}) might define the geodesics of the de Sitter spacetime. 

\section{de Sitter geodesics from a variational principle} 

Let us consider a particle of mass $m$, whose action functional is given by
\be
S = -\, m c \int_a^b ds,
\label{graviaction}
\ee
with $ds = (g_{\alpha \beta} \, dx^\alpha dx^\beta)^{1/2}$. Then comes the crucial point: in order to consider all possible trajectories between points $a$ and $b$, the variational principle must take into account the transitivity properties of the underlying spacetime. In the case of de Sitter, this can be done by considering the spacetime variations
\be
\delta_{\Pi} x^\mu = \xi^{\mu}_{\rho} \, \delta x^{\rho},
\ee
with $\delta x^{\rho}$ an ordinary spacetime variation. Accordingly, the action variation reads
\begin{equation}
\delta S = - mc \int_{b}^{a} \left[ \onehalf \, \partial_\gamma (g_{\alpha\beta}) \, u^{\alpha} dx^{\beta} \, \delta_{\Pi}x^{\gamma} +
g_{\alpha\beta} \, u^{\alpha} \, \delta_{\Pi} (dx^{\beta}) \right],
\label{eq:2}
\end{equation}
with $u^\alpha = d x^\alpha/ds$ the particle four-velocity. Using the trivial identity $\delta_{\Pi} (dx^{\beta}) = d (\delta_{\Pi} x^{\beta})$ in the last term, we get
\begin{equation}
\delta S = - mc \int_{b}^{a} \left[ \onehalf \, \partial_\gamma (g_{\alpha\beta}) \, u^{\alpha} dx^{\beta} \, \delta_{\Pi}x^{\gamma} +
g_{\alpha\beta} \, u^{\alpha} \, d (\delta_{\Pi} x^{\beta}) \right].
\label{eq:2bis}
\end{equation}
Integrating the last term by parts, and neglecting the surface term, the variation becomes
\begin{equation}
\delta S = - mc \int_{b}^{a} \left[ \frac{1}{2} \frac{\partial g_{\alpha\beta}}{\partial x^{\gamma}} \, u^{\alpha} u^{\beta} -
\frac{d}{ds} \left( g_{\alpha\gamma} \, u^{\alpha} \right)  \right]
\xi_{\rho}^{\gamma} \, \delta x^\rho \, ds.
\label{eq:3}
\end{equation}
After some algebraic manipulation, it can be rewritten in the form
\be
\delta S = m c \int_{b}^{a} \left[g_{\alpha\beta} \, u^{\gamma}
\nabla_{\gamma} u^{\alpha} \, \xi_{\rho}^{\beta} \right] \delta x^{\rho} \, ds,
\ee
with $\nabla_\gamma$ a covariant derivative in the Christoffel connection of the metric $g_{\alpha\beta}$. Considering that $\xi_{\rho}^{\beta}$ satisfy the Killing equation, we can write
\be
\delta S = \int_{b}^{a} \left[ u^{\gamma}
\nabla_{\gamma} (\xi_{\rho}^{\beta} \, p_{\beta}) \right] \delta x^{\rho} \, ds.
\ee
From the invariance of the action, and taking into account the arbitrariness of $\delta x^{\rho}$, we obtain
\be
\frac{d \pi_\rho}{ds} -
\Gamma^\nu{}_{\rho \gamma} \, \pi_\nu \, u^\gamma = 0, 
\label{GendSGeod}
\ee
which is just the geodesic equation (\ref{GendSGeod0}). Owing to the fact that these geodesics are consistent with the transitivity properties of the de Sitter spacetime, any two points of this space will be connected by a geodesic of this family. For this reason, they can be considered to be the true geodesics of the de Sitter spacetime.

Although obtained in stereographic coordinates, due to the fact that the geodesic equation is covariant, it will have the same form in any other coordinate system. Of course, the explicit form of the Killing vectors $\xi_{\rho}^{\beta}$, and consequently of the de Sitter momentum $\pi^\mu$, will be different in different coordinates; the geodesic equation, however, remains formally the same. The basic advantage of the stereographic coordinates is that the separation of the conserved current into ordinary and proper conformal momenta becomes manifest.

\section{Final remarks}

Both Minkowski and de Sitter are fundamental spacetimes in the sense that, as quotient spaces, they are known {\it a priori}, independently of Einstein equation. Both are isotropic and homogeneous, but their homogeneity properties differ substantially: whereas Minkowski is transitive under translations, de Sitter is transitive under a combination of translations and proper conformal transformations. Now, transitivity is intimately related to the notion of motion. For example, any two points of Minkowski spacetime are connected by a spacetime translation. As a consequence, motion in this spacetime is described by trajectories whose points are connected to each other by ordinary translations. On the other hand, any two points of de Sitter spacetime are connected to each other by a combination of translation and proper conformal transformations---the so-called de Sitter ``translations''. As a consequence, the notion of motion in this spacetime will change in the sense that it will be described by trajectories whose points are connected to each other by a combination of translation and proper conformal transformations. However, in the usual procedure to obtain the geodesics of the de Sitter spacetime, given by
\be
\frac{d u_\rho}{ds} -
\Gamma^\nu{}_{\rho \gamma} \, u_\nu \, u^\gamma = 0, 
\label{dSGeod}
\ee
the appropriate homogeneity properties are not taken into account in the variational principle. As a consequence, there are points in de Sitter spacetime which are not connected by any one of these geodesics \cite{ellis}. This single fact constitutes a clear evidence that they do not represent the true geodesics of the de Sitter spacetime.

On the other hand, by taking into account the appropriate transitivity properties of the de Sitter spacetime, we have obtained a new family of trajectories, given by
\be
\frac{d U_\rho}{ds} -
\Gamma^\nu{}_{\rho \gamma} \, U_\nu \, u^\gamma = 0, 
\label{GendSGeodBis}
\ee
where
\be
U_\rho = \xi_{\rho}^{\beta} \, u_{\beta} \equiv
\big[ \delta_\rho^\beta - \frac{1}{4l^2} \, (2 \eta_{\rho \nu} \, x^\nu x^\beta -
\sigma^2 \delta_\rho^\beta)\big] u_\beta
\ee
is an anholonomic four-velocity, which takes into account the translational and the proper conformal `directions' of the de Sitter spacetime. As a consequence, the corresponding trajectories include both notions of motion: translational and proper conformal. They are, consequently, able to connect any two points of de Sitter spacetime. Furthermore, similarly to what happens in ordinary special relativity, these trajectories coincide with the conservation of momentum, which in the de Sitter case includes, in addition to ordinary momentum, also the proper conformal momentum. They are consistent, therefore, with the de Sitter kinematics. Since the expression describing these trajectories can be obtained from a variational principle --- and are for this reason maximizing curves --- they can be interpreted as the true geodesics of the de Sitter spacetime. The only difference between equations (\ref{dSGeod}) and (\ref{GendSGeodBis}) is that in the de Sitter case the particle four-velocity is anholonomic.

The de Sitter geodesics give rise to a new concept of motion, defined as a combination of translation and proper conformal transformation. This kind of motion holds not only in de Sitter spacetime, but in any gravitationally-curved generalization that reduces locally to de Sitter. The relative importance between the two involved notions of motion is determined by the value of the de Sitter pseudo-radius $l$. For large values of $l$ in relation to the Planck length $l_P$, motion will be preponderantly determined by spacetime translations. In the formal limit $l \to \infty$, de Sitter contracts to Minkowski, and motion will be determined by ordinary translations only. In this case, the conformal degrees of freedom are turned off, and ordinary momentum turns out to be conserved. On the other hand, for values of $l$ of the order of the Planck length $l_P$, motion will be preponderantly determined by proper conformal transformations.\footnote{In the formal limit $l \to 0$, the de Sitter group contracts to the conformal Poincar\'e group --- given by the semi-direct product between Lorentz and the proper conformal transformations \cite{zimer} --- and the de Sitter spacetime reduces to a cone spacetime, which is transitive under proper conformal transformations \cite{cone}. It is important to reinforce that $l \to 0$ is just a formal limit. The output of this limit should be thought of as the `frozen geometrical structure' behind the spacetime quantum fluctuations taking place at the Planck scale.} 
If spacetime kinematics is to be ruled by the de Sitter group \cite{new} --- as it should, for example, in the presence of a cosmological constant --- this new notion of motion may possibly have important consequences for the physics of the Planck scale, where conformal invariance plays a fundamental role.

\section*{Acknowledgements}
The authors would like to thank FAPESP, CAPES and CNPq for financial support.


\end{document}